\documentclass[prl,twocolumn,amssymb]{revtex4}
\usepackage{graphicx}
\usepackage{ulem}

\begin{document}

\title{Degeneracy of Many-body Quantum States in an Optical Lattice under a Uniform Magnetic Field }
\author{Jian Zhang$^{1}$, Chao-Ming Jian$^{1}$, Fei Ye$^{2}$ and Hui Zhai$^{1}$}
\affiliation{$^1$ Institute for Advanced Study, Tsinghua University, Beijing, 100084, P. R. China\\
$^2$ College of Material Science and Optoelectronics Technology, Graduated University of Chinese Academy of Science,  Beijing 100049, P. R. China}
\date{\today}

\begin{abstract}

We prove a theorem that shows the degeneracy of many-body states for particles in a periodic lattice and under a uniform magnetic field depends on the total particle number and the flux filling ratio. Non-interacting fermions and weakly interacting bosons are given as two examples. For the latter case, the phenomenon can also be physically understood in terms of destructive quantum interference of multiple symmetry-related tunneling paths between classical energy minima, which is reminiscent of the spin-parity effect discovered in magnetic molecular clusters. We also show that the quantum ground state of a mesoscopic number of bosons in this system is not a simple mean-field state but a fragmented state even for very weak interactions.

\end{abstract}

\maketitle

Recently, cold atoms in optical lattices subject to a large effective magnetic field has become a new direction in cold atom research. Interacting bosons in rotating optical lattices have been investigated experimentally \cite{exp}, where rotation plays a similar role to magnetic field for charged particles. A synthetic magnetic field for neutral atoms has also been successfully realized through engineering atom-light interaction, and this technique can be applied to optical lattices straightforwardly \cite{speilman}. These experimental progresses and the rich physics in such a system have generated lots of theoretical interest \cite{theorymf,MC,Cooper,Zhai}. Most of the work so far are mean-field studies or classical Monte Carlo simulations of mean-field states \cite{theorymf,MC,Zhai}, and some focus on strongly correlated quantum-Hall states in the strong interaction limit \cite{Cooper}. In this letter, (i) we shall first prove a general theorem on the degeneracy of many-body states in this system, and (ii) we shall present non-interacting fermions and weakly interacting bosons as two concrete examples of this theorem. In addition, (iii) we will show that in the case of finite number of bosons, the system exhibits an intriguing non-mean-field ground state, even in the regime of very weak interactions.

We consider a two-dimensional optical lattices and a uniform effective magnetic field $B$ along $\hat{z}$ direction (created by rotation or atom-light interaction). In this work, we focus on the properties of a uniform system, and do not consider the harmonic trapping potential. Under the Landau gauge, the single particle Hamiltonian is given by
\begin{equation}
\hat{H}_0=-\frac{\hbar^2}{2m}\partial^2_x+\frac{1}{2m}\left(-i\hbar\partial_y-eBx\right)^2+V_{\text{OL}}(x,y).\label{singleH}
\end{equation}
${\bf R_1}$ and ${\bf R_2}$ are the basis vectors of the optical lattice potential $V_{\text{OL}}(x,y)$.
The Hamiltonian for a many-body system reads
\begin{equation}
\hat{H}=\sum\limits^{N}_{i=1}\hat{H}_0({\bf r}_i)+\sum\limits_{i<j}V({\bf r_i}-{\bf r_j}).
\end{equation}
Let $N$ denote the total number of particles, and $\nu=\phi/\phi_0$ denote the flux filling ratio, where $\phi=B({\bf R_1}\times{\bf R_2})\cdot\hat{z}$ is the effective magnetic flux per plaquette and $\phi_0=h/e$ is the flux quantum.

{\bf Theorem on Degeneracy.} {\it For $\nu=p/q$ where $p$ and $q$ are coprime numbers, $q$ is a prime number, and $N/q$ is not an integer, all many-body eigenstates are at least $q$-fold degenerate.}

To prove this theorem, we first introduce the magnetic translation operator for particle $j$ as $\hat{T}_j({\bf r})=\exp\{i{\bf r}{\bf \Pi}_j/\hbar\}$,
where $\Pi_{jx}=-i\hbar\partial_{x_j}-eB y_j$ and $\Pi_{jy}=-i\hbar\partial_{y_j}$. One can show that both $\hat{T}_j({\bf R_1})$ and $\hat{T}_j({\bf R_2})$ commute with $\hat{H}_0({\bf r_j})$, but
\begin{equation}
\hat{T}_j({\bf R_1})\hat{T}_j({\bf R_2})=e^{i2\pi p/q}\hat{T}_j({\bf R_2})\hat{T}_j({\bf R_1}).
\end{equation}
Furthermore we define the magnetic translation operator for all $N$ particles together of one basis vector as
\begin{equation}
\hat{T}_X=\prod_j \hat{T}_j({\bf R_1}), \   \  \hat{T}_Y=\prod_j  \hat{T}_j({\bf R_2}).
\end{equation}
Both $\hat{T}_X$ and $\hat{T}_Y$ commute with $\sum_{i<j}V({\bf r_i-r_j})$ since all particles are translated together. Therefore, we have
\begin{equation}
[\hat{T}_{X},\hat{H}]=[\hat{T}_Y,\hat{H}]=0.
\end{equation}
However,
\begin{equation}
\hat{T}_X\hat{T}_Y=e^{i2\pi p N/q}\hat{T}_Y\hat{T}_X.
\end{equation}
We can choose a common eigenstate of $\hat{H}$ and $\hat{T}_Y$ as $\Psi_0$ with $\hat{T}_Y\Psi_0=\eta \Psi_0$ and $\hat{H}\Psi_0=\epsilon \Psi_0$. Defining $\Psi_l=(\hat{T}_X)^l\Psi_0$, $(l=1,\dots,q-1)$, then
\begin{eqnarray}
\hat{T}_Y\Psi_l&=&e^{-2\pi i p lN/q} \eta \Psi_l,\\
\hat{H}\Psi_l&=&\epsilon \Psi_l
\end{eqnarray}
If $p$ and $q$ are coprime numbers, and $q$ is a prime number, when $N/q$ is not an integer, $pNl$ with $l=0,1,\dots, q-1$ are different integers modulo $q$, therefore all $\Psi_l$ have different eigenvalues of $T_Y$, i.e., they are orthogonal states with the same energy.

This theorem adds one more example to the rare cases that an exact theorem can be proved for a many-body system. It holds for any pair-wise interaction $V({\bf r})$, for both bosonic and fermionic systems, and for all lattice geometry such as rectangular, triangular or hexagonal. It can also be generalized to a three-dimensional cubic lattice straightforwardly, since both $T_X$ and $T_Y$ commute with translations along the $\hat{z}$-direction. This theorem imposes a strong constraint on all the many-body theory of this system, namely, any approximation scheme applied to the system must respect the theorem.  Hereafter, we shall give two concrete examples using the two dimensional square lattice (${\bf R_1}=a\hat{x}$ and ${\bf R_2}=a\hat{y}$), from which we hope to provide a physical picture for this phenomenon.

The first example is non-interacting fermions. For $\nu=p/q$, the eigenstates of $H_0$ satisfy the magnetic Bloch theorem and are characterized by two good quantum numbers $k_x\subset (-\pi/(aq), \pi/(aq)]$ and $k_y\subset (-\pi/a,\pi/a]$ \cite{TKNN}. Due to the magnetic translation symmetry, each single particle state is $q$-fold degenerate \cite{deg,Zhai}. Therefore, the ground state can not be unique unless $N$ is a multiple of $q$. While if $N$ modulo $q$ is $l$, the degeneracy is at least $q!/(l!(q-l)!)$, which is a multiple of $q$.

The second example is weakly interacting bosons. Taking $\nu=1/3$ as an example, with a simplified effective model, we find that the ground state is non-degenerate if $N$ is a multiple of three, while three-fold degenerate otherwise. This is reminiscent of the spin-parity effect discovered in the studies of spin coherence of magnetic molecular clusters \cite{spin-parity}, where distinctive behaviors are found in spin integer and half-interger systems. There, the phenomenon is interpreted in terms of the quantum interference of multiple symmetry-related tunneling paths connecting the degenerate classical states. By similar quantum interference argument, we can understand the degeneracy for $\nu=1/3$ and more general cases where no analogy of spin-parity effect has been discussed before.

{\it Effective Model:} Since each single particle state is $q$-fold degenerate, there must be $q$-fold degenerate single particle ground states. For $\nu=1/3$, three degenerate single particle ground states $\varphi_{1,2,3}({\bf r})$ are magnetic Bloch states with $(k_x,k_y)=(0,0)$, $(0,2\pi/(3a))$ and $(0,-2\pi/(3a))$. Let $\hat{b}_{1,2,3}$ be boson operators for these three modes, then all the states $\hat{b}_1^{\dag n_1}\hat{b}_2^{\dag n_2}\hat{b}^{\dag (N-n_1-n_2)}_3|0\rangle$ ($n_1=0,\dots, N$, $n_2=0,\dots, N-n_1$) are degenerate without interactions. Interactions will result in two effects: it will mix these $(N+1)(N+2)/2$ states, and also introduce quantum depletion to other single-particle excited states. In the regime that the interaction strength is much smaller than the magnetic band width, the first effect is dominant due to the boson enhancement factor and the absence of single-particle energy cost. In practice, the band width is usually a small fraction ($\sim 0.01$) of $\hbar^2/(2ma^2)$, while the interaction $\sim \hbar^2 a_{\text{s}}/(ma^3)$, hence it requires that the scattering length $a_{\text{s}}$ should be at least three order of magnitude smaller than $a \sim 0.5 \mu m$. Atoms like $^{39}$K, $^7$Li and $^{133}$Cs, which either have very small background scattering length or have the scattering length that can be tuned across zero by a Feshbach resonance, are particularly suitable for reaching this limit. 

Hence, we take the approximation that only these three modes are kept, then the single particle term becomes a constant. Taking $V({\bf r})=U\delta({\bf r})$, an effective Hamiltonian is purely given by the interaction as
\begin{eqnarray}
\hat{H}_{\text{eff}}=\alpha(\hat{n}^2_1+\hat{n}^2_2+\hat{n}^2_3)+4\beta(\hat{n}_1\hat{n}_2+\hat{n}_2\hat{n}_3+\hat{n}_1\hat{n}_3)\nonumber\\
+[2\gamma (\hat{b}^\dag_1\hat{b}^\dag_1\hat{b}_2\hat{b}_3+\hat{b}^\dag_2\hat{b}^\dag_2\hat{b}_1\hat{b}_3+\hat{b}^\dag_3\hat{b}^\dag_3\hat{b}_1\hat{b}_2)+\text{h.c.}],\label{H13}
\end{eqnarray}
where $\alpha=U\int |\varphi_{1}|^4d^2{\bf r}$, $\beta=U\int |\varphi_{1}|^2|\varphi_{2}|^2d^2{\bf r}$ and $\gamma=U\int \varphi^{*2}_{1}\varphi_2\varphi_3 d^2{\bf r}$. A $90^\circ$ rotational symmetry leads to $\alpha-2\beta=2\gamma$. This model contains all the four-boson interaction terms allowed by momentum conservation.

\begin{figure}[tbp]
\includegraphics[height=1.6in, width=3.4 in]
{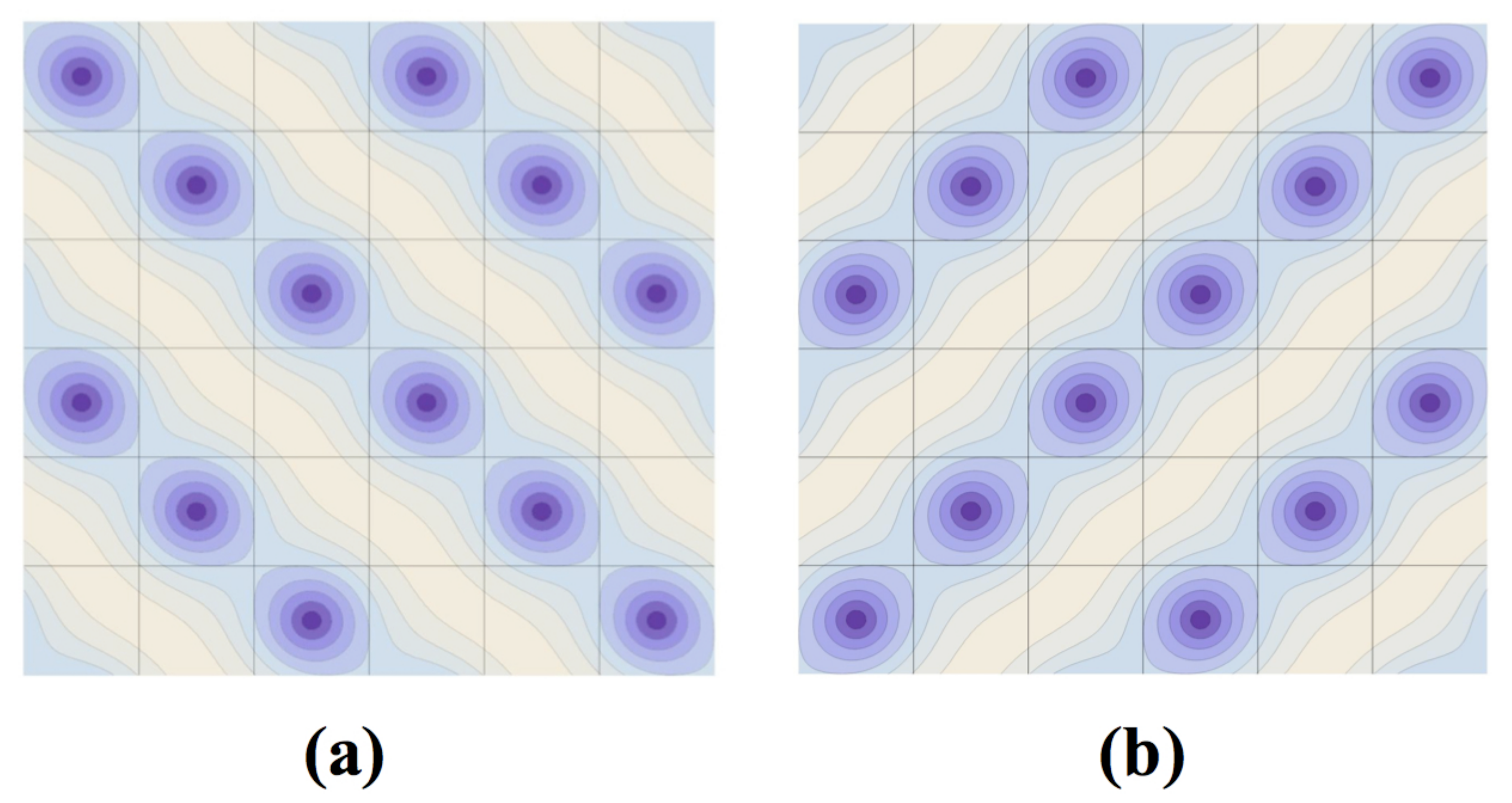}
\caption{(Color online) This is a real space contour plot for the amplitude of the condensate wave function $(\varphi_1({\bf r})+\varphi_2({\bf r})+e^{i2\pi/3}\varphi_3({\bf r}))/\sqrt{3}$ (a) and $(\varphi_1({\bf r})+\varphi_2({\bf r})+e^{-i2\pi/3}\varphi_3({\bf r}))/\sqrt{3}$ (b).  The intersections of the vertical and the
horizontal lines indicate the lattice sites, i.e. the potential minima of $V_{\text{OL}}(x,y)$. The dark area indicates the locations of the vortices. \label{vortex}}
\end{figure}

{\it Mean-field Analysis.} We first revisit the mean-field states (MFS) in the weakly interacting regime. Previous mean-field studies of this system are limited to the Gutzwiller mean-field ansatz in tight-binding limit \cite{theorymf}. Here we implement another MFS written in the magnetic Bloch state bases, which are
\begin{equation}
\frac{1}{\sqrt{N!}}\left(u_1 \hat{b}^\dag_1+u_2 \hat{b}^\dag_2+u_3 \hat{b}^\dag_3\right)^N|0\rangle,
\end{equation}
By minimizing the energy we find six degenerate MFS. Three of them are $\Psi_j=\hat{a}_j^{\dag N}|0\rangle/\sqrt{N!}$ ($j=1,2,3$), where $\hat{a}_1=(\hat{b}_1+\hat{b}_2+e^{i2\pi/3}\hat{b}_3)/\sqrt{3}$, $\hat{a_2}$ and $\hat{a}_3$ follow by cyclic permutation of the coefficients in $\hat{a}_1$. The other three are $\Phi_j=d_j^{\dag N}|0\rangle/\sqrt{N!}$ ($j=1,2,3$), where $\hat{d}_1=(\hat{b}_1+\hat{b}_2+e^{-i2\pi/3}\hat{b}_3)/\sqrt{3}$, $\hat{d_2}$ and $\hat{d}_3$ follow by cyclic permutation of the coefficients in $\hat{d}_1$.

The vortex lattice configurations are displayed in Fig. \ref{vortex}. Here we obtain the magnetic Bloch function $\varphi_i({\bf r})$ by diagonalizing the lattice potential $V_{\text{OL}}$ in the lowest Landau level subspace \cite{TKNN}. These results agree with the Gutzwiller mean-field studies \cite{theorymf} and the classical Monte Carlo simulation in the tight-binding limit \cite{MC}. As one can see from Fig. \ref{vortex}, the unit cell is enlarged to $q\times q$ with vortex lattices, and therefore the symmetry of the vortex lattice state is lower than that of the original Hamiltonian. Hence, the MFS are degenerate. The degeneracy of the MFS is also a manifestation of our theorem in the thermodynamic limit, and this degeneracy is crucial for the later discussion of the quantum ground state.

\begin{figure}[tbp]
\includegraphics[width=3.4in,height=1.4in]
{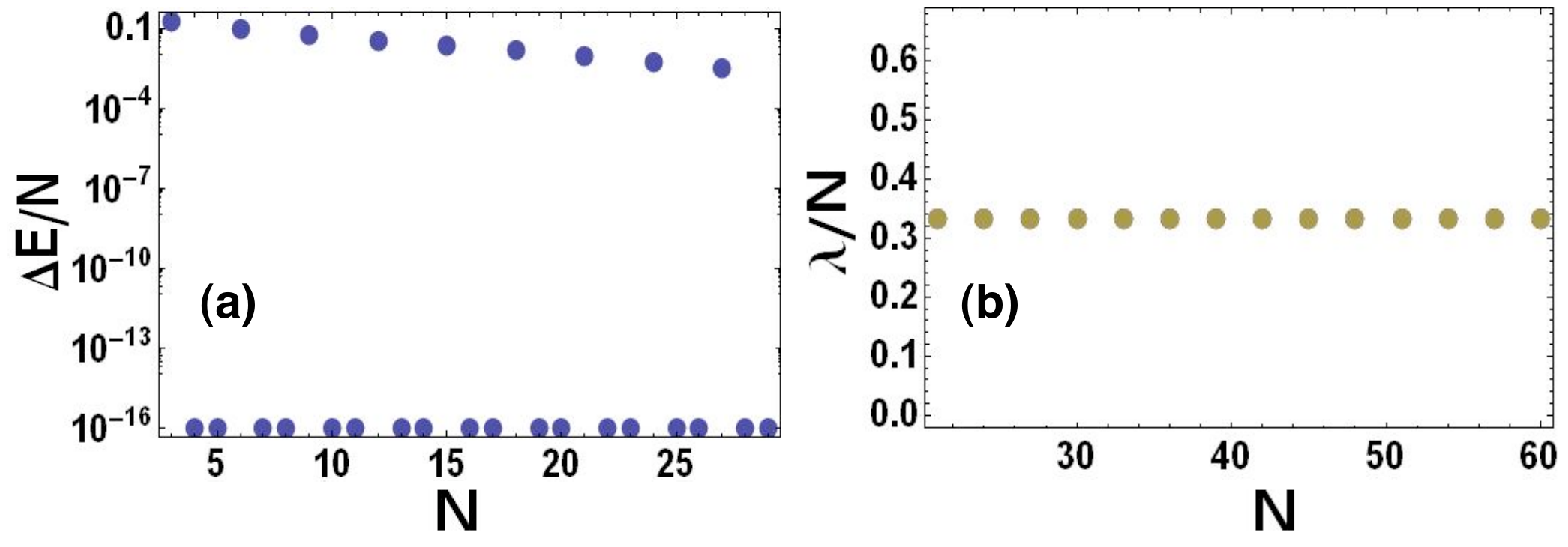}
\caption{(a) $\Delta E=(E_1-E_0)/N$ and $(E_2-E_0)/N$ as a function of $N$. $E$ is in unit of $U$. (b): the eigenvalues of $\lambda_i/N$ of the density matrix $\hat{\rho}$ of the ground state.
\label{ED}}
\end{figure}

\begin{figure}[tbp]
\includegraphics[width=3.4in]{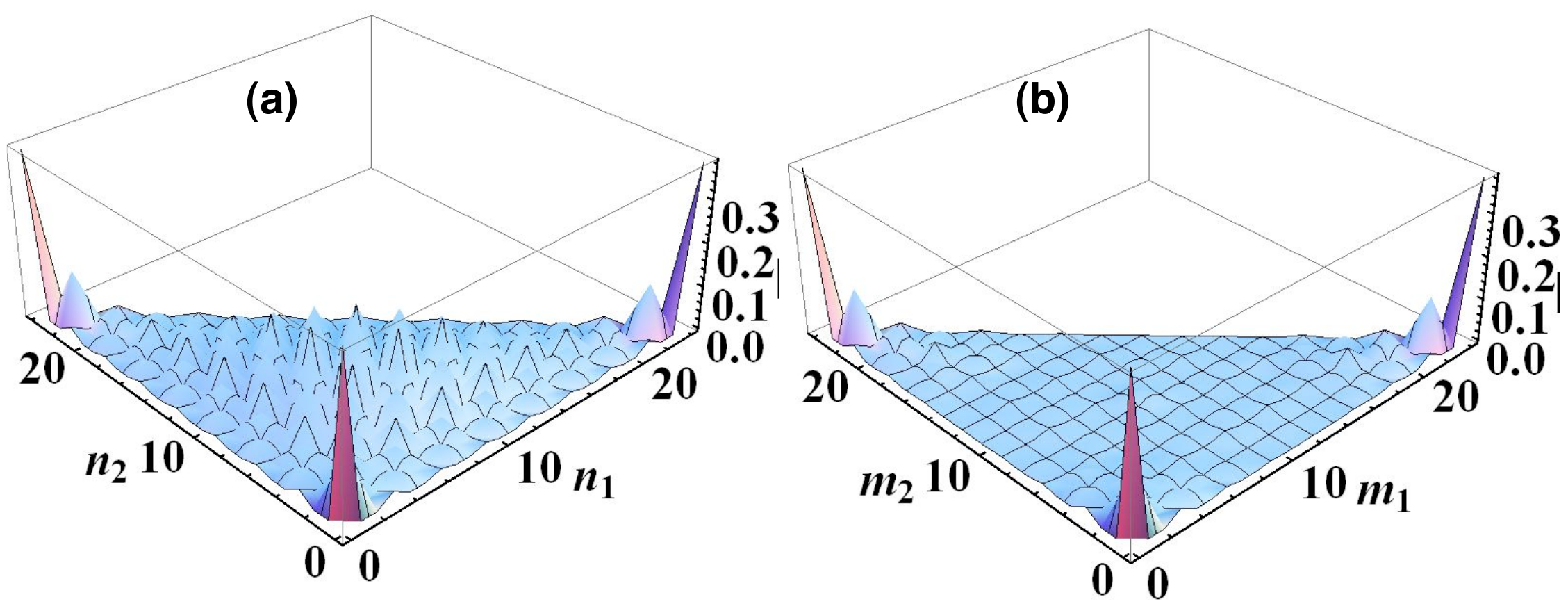} \caption{(a) $|\mathcal{A}_{n_1,n_2}|$ as a function of $(n_1,n_2)$ (b) $|\mathcal{D}_{m_1,m_2}|$ as a function of $(m_1,m_2)$. See the definition in the text. Here $N=24$.\label{cat} }
\end{figure}

{\it Exact Diagonalization Studies.} Given the total number of bosons $N$, we diagonalize the Hamiltonian Eq. \ref{H13} using the Fock state bases. We sort the eigen-energies from the lowest to the highest as $E_n$. For the ground state, we compute the density matrix $\hat{\rho}=\langle \hat{b}^\dag_i \hat{b}_j\rangle$ and its eigenvalues $\lambda_i$. The main results from this exact diagonalization are presented in Fig. \ref{ED} and \ref{cat}.

In Fig. \ref{ED}(a) we plot $\Delta E=E_1-E_0$ and $E_2-E_0$ for different $N$, which shows that  the ground state is three-fold degenerate for $N=3n+1$ and $3n+2$, while it is not degenerate for $N=3n$. Nevertheless, it also shows that $\Delta E$ decreases exponentially as $N$ increases. As we will show later, $\Delta E$ is in fact the tunneling splitting due to macroscopic quantum tunneling between different MFS with the same classical energy.  In the limit $N\rightarrow +\infty$ the difference vanishes and all states are practically at least $q$-fold degenerate. This numerical calculation clearly verifies our theorem.

In addition, in the case of a non-degenerate ground state, we show in Fig. \ref{ED}(b) that all three eigenvalues $\lambda_i$ equal to $N/3$. Hence, the non-degenerate ground state is a fragmented state, instead of a single condensate. Taking $\nu=1/3$ and $N=24$ as an example, the structure of the quantum ground state is further illustrated in Fig. \ref{cat}. We first rotate the single-particle bases to $\{\hat{a}_j (j=1,2,3)\}$. The wave function is written as
\begin{equation}
\sum_{n_1,n_2}\mathcal{A}_{n_1,n_2}\frac{\hat{a}^{\dag n_1}_1\hat{a}^{\dag n_2}_2\hat{a}^{\dag N-n_1-n_2}_{3}}{\sqrt{n_1! n_2! (N-n_1-n_2)!}}|0\rangle
\end{equation}
$|\mathcal{A}_{n_1,n_2}|$ is plotted in Fig. \ref{cat}(a), where the most pronounced feature is the three peaks located at $(n_1,n_2)=(N,0)$, $(0,N)$ and $(0,0)$. It means that the ground state contains an equal weight superposition of three MFS $\Psi_j$ ($j=1,2,3$). Another feature in Fig. \ref{cat}(a) is that there is also weight of $\mathcal{A}_{n_1,n_2}$ distributed broadly in the center region in the plot. This part of wave function can be recast as
\begin{equation}
\sum_{m_1,m_2}\mathcal{D}_{m_1,m_2}\frac{\hat{d}^{\dag m_1}_1\hat{d}^{\dag m_2}_2\hat{d}^{\dag N-m_1-m_2}_{3}}{\sqrt{m_1! m_2! (N-m_1-m_2)!}}|0\rangle.
\end{equation}
$|\mathcal{D}_{m_1,m_2}|$ is plotted in Fig. \ref{cat}(b), which again shows three sharp peaks located at $(m_1,m_2)=(N,0)$, $(0,N)$ and $(0,0)$. Fig. \ref{cat} indicates the quantum ground state of this case can be well approximated (with overlap $\gtrsim 90\%$) as a coherent superposition of six degenerate MFS with appropriate coefficients. For a general situation of $\nu=1/3$, the detailed structures of the ground state are found from numerical calculations and are illustrated in Table I.

\begin{table}
\begin{ruledtabular}
\begin{tabular}{|c|c|c|}
N & D & Wave functions \\
\hline
 $3n$ & 1 & $\Psi_1+\Psi_2+\Psi_3+e^{i\pi N/6}(\Phi_1+\Phi_2+\Phi_3)$ \\
 \hline
$3n+1$&  &  $e^{-i2\pi N/3}\Psi_1+\Psi_2+\Psi_3+\sqrt{3}e^{-i\pi N/2}\Phi_1$,   \\
 or  &  3  & $\Psi_1+e^{-i2\pi N/3}\Psi_2+\Psi_3+\sqrt{3}e^{-i\pi N/2}\Phi_2$, \\
 $\  3n+2$& &   $\Psi_1+\Psi_2+e^{-i2\pi N/3}\Psi_3+\sqrt{3}e^{-i\pi N/2}\Phi_3$
\end{tabular}
\caption{For $\nu=1/3$, the relation between $N$, the degeneracy (D) and the ground state wave functions. \label{table}}
\end{ruledtabular}
\end{table}

Such a fragmented state results from quantum tunneling between degenerate classical MFS. During the past years, quite a few of such examples have been studied in cold atoms systems. Fragmentation has been studied for bosons in double-well \cite{Hocat}, multi-well \cite{multiwell}, and also in a single trap \cite{dipole} or in a triple-well with dipolar interactions \cite{dipolethree}. The exact ground state of spin-1 bosons in the ``polar" phase is found to be a singlet-pair condensate \cite{LPB}. Breakdown of mean-field approximation and fragmented ground states have also been found in a rotating BEC at critical frequency of vortex nucleation \cite{nucleation,burnett} and in the fast rotating regime of bosons \cite{rotating}. These states, in principle, can be distinguished from MFS through measurement of noise statistics and interference pattern \cite{Hocat}. Nevertheless, they are usually very fragile and is difficult to prepare, in particular, for large $N$ when the tunneling splitting between different MFS becomes exponentially small. However, the decay rate of this tunneling splitting depends on the details of each model. It is still possible that there exists a regime that the decay rate is sufficiently low and a fragmented state with a good number of particles can be achieved experimentally. We leave this for future studies.

{\it Quantum Interference Viewpoint of Degeneracy}. Previous studies of spin coherence in magnetic molecular clusters have revealed an effect known as the spin-parity effect  \cite{spin-parity}. Consider a $SU(2)$ spin model with time-reversal symmetry, and two classical energy minima located at the south and north poles of the Bloch sphere. One can show that for each tunneling path between these two classical minima, there is always another path related by time-reversal symmetry, and therefore these two paths have equal tunneling amplitudes. One can also show that there is a relative phase $\exp\{i 2\pi S\}$ between the two paths. Therefore, the total tunneling amplitude is proportional to $1+\exp\{i 2\pi S\}$, which vanishes for half-integer $S$ and results in double degeneracy \cite{spin-parity}. A two-mode boson model can be mapped to a $SU(2)$ spin model with $S=N/2$. Recently, a similar effect has also been discussed for BEC in double-well \cite{rong} and should also exist in the model discussed in Ref. \cite{dipole}. By the Schwinger boson scheme, $H_{\text{eff}}$ of Eq. \ref{H13} is mapped to a $SU(3)$ spin model. However, to the best of our knowledge, no such effect has been reported in a $SU(N)$ spin model for $N\geqslant 3$ before. The discussion below is equivalent to a generalized spin-parity effect for a SU(3) spin model, and similar analysis can even be possibly applied to a general $SU(N)$ case.

Using the coherent state representation, we write
\begin{equation}
\Psi_j
=\frac{1}{\pi}\int d^2z_j \frac{\bar{z_j}^N}{\sqrt{N!}}e^{-|z_j|^2/2}|z_j\rangle .\label{coherent-state}
\end{equation}
where $|z_j\rangle=e^{-|z_j|^2/2}e^{z_j \hat{a}^\dag_j}|0\rangle$.
The tunneling amplitude between $\Psi_1$ and $\Psi_2$ can be formulated in terms of the imaginary time coherent state path integral as \cite{spin-parity}
\begin{eqnarray}
&U_{12}&=\langle \Psi_1|e^{-\hat{H}\tau/\hbar}|\Psi_2\rangle=\int\mathcal{D}\Omega\times \nonumber\\
& & \int d^2z_1 d^2z_2\frac{(z_1\bar{z}_2)^N}{\pi^2 N!}e^{-(|z_1|^2+|z_2|^2)/2}e^{-S[\Omega]/\hbar}\label{Integral}
\end{eqnarray}
where $S[\Omega]$ is the Euclidean action for path $\Omega$ from the initial coherent state $|z_1\rangle$ to the final one $|z_2\rangle$. We note that
$H_{\frac{1}{3}}$ is invariant under two different magnetic translation operations:
\begin{eqnarray}
&&(\text{i}) \   \   \hat{b}_1\rightarrow \hat{b}_3e^{-i2\pi/3}, \  \  \hat{b}_2\rightarrow \hat{b}_1e^{i2\pi/3}, \   \ \hat{b}_3\rightarrow \hat{b}_2; \nonumber\\
&&(\text{ii}) \  \   \hat{b}_1\rightarrow \hat{b}_2e^{-i2\pi/3}, \  \  \hat{b}_2\rightarrow \hat{b}_3, \  \ \hat{b}_3\rightarrow \hat{b}_1e^{i2\pi/3},\nonumber
\end{eqnarray}
under which $\hat{a}_1\rightarrow e^{\pm i 2\pi/3}\hat{a}_1$, $\hat{a}_2\rightarrow e^{\pm i 2\pi/3}\hat{a}_2$, and $\hat{a}_3\rightarrow \hat{a}_3$ (upper signs for (i) and lower signs for (ii)). Hence, for any tunneling path $\Omega^0$ (connecting initial state $|z^0_1\rangle$ and final state $|z^0_2\rangle$), there must be two other symmetry-related paths $\Omega^1$ and $\Omega^2$ (connecting $|z^i_1\rangle$ and $|z^i_2\rangle$ $(i=1,2)$, respectively), which satisfy $S[\Omega^2]=S[\Omega^1]=S[\Omega^0]$ and $z^{1,2}_{1}=e^{\pm i2\pi/3}z^{0}_{1}$ and $z^{1,2}_{2}=e^{\mp i2\pi/3}z^{0}_{2}$. By Eq. \ref{Integral}, one has $U^1_{12}=e^{i 2\pi N/3}U^0_{12}$ and  $U^2_{12}=e^{-i2\pi N/3}U^0_{12}$, where $U^i_{12}$ ($i=0,1,2$) denotes the tunneling amplitudes of path $\Omega^i$. Thus, the total tunneling amplitude is always proportional to $1+e^{i 2\pi N/3}+e^{-i 2\pi N/3}$. For $N=3n+1$, $3n+2$, this tunneling vanishes due to destructive interference. This is a generalization of $1+e^{i 2\pi S}$ ($S=N/2$) for the spin-parity effect. Similarly, one can show the tunneling between all $\Psi_i$ and $\Psi_j$ ($i\neq j$), and the tunneling between $\Phi_i$ and $\Phi_j$ ($i\neq j$) exhibits the same behavior. Hence, we have established an alternative viewpoint from quantum interference as to why the case $N$ being a multiply of three is different.

{\it Acknowledgment}: We thank Tin-Lun Ho, Mehmet Oktel, Zhan Xu, Uwe Fischer, Xiao-Gang Wen for helpful discussions. HZ is supported
by the Basic Research Young Scholars Program of Tsinghua
University, NSFC Grant No. 10944002, and FY is supported by NSFC Grant No. 10904081.

\end{document}